\documentclass[prb,twocolumn]{revtex4}%

\newcommand{\be}{\begin{equation}}
\newcommand{\ee}{\end{equation}}
\newcommand{\bea}{\begin{eqnarray*}}
\newcommand{\eea}{\end{eqnarray*}}
\newcommand{\bean}{\begin{eqnarray}}
\newcommand{\eean}{\end{eqnarray}}

\begin{document}

\draft
\title
{\bf  Thermal rectification properties of multiple-quantum-dot junctions}

\author{ David M.-T. Kuo$^{1}$ and Yia-chung Chang$^{2}$  }

\address{$^{1}$Department of Electrical Engineering and Department of Physics, National Central
University, Chungli, 320 Taiwan}

\address{$^{2}$Research Center for Applied Sciences, Academic
Sinica, Taipei, 115 Taiwan}

\date{\today}

\begin{abstract}
It is illustrated that semiconductor quantum dots (QDs)
embedded into an insulating matrix connected with metallic
electrodes and some vacuum space can lead to significant thermal rectification effect.
A multilevel Anderson model is used to investigate the thermal rectification properties of
the multiple-QD junction. The charge and heat currents in the
tunneling process are calculated via the Keldysh Green's
function technique. We show that pronounced thermal
rectification and negative differential thermal conductance (NDTC)
behaviors can be observed for the multiple-QD junction with asymmetrical
tunneling rates and strong interdot Coulomb interactions.
\end{abstract}

\maketitle

 Records of thermal rectification date back to 1935 when Starr discovered that
copper oxide/copper junctions can display a thermal diode
behavior.$^{1}$ Recently, thermal rectification effects have been
predicted to occur in one dimensional phonon junction
systems.$^{2-5}$ Such a thermal rectification effect is crucial
for heat storage. Scheibner and coworkers have experimentally
observed the asymmetrical thermal power of the two-dimensional
electron gas in QD under high magnetic fields.$^{6}$ So far the
rectification mechanism of a single QD is still ambiguous owing to
the unclear relation between the thermal power and the thermal
rectification effect. This inspires us to investigate whether the
QD junction system can act as a thermal rectifier. A useful
thermal diode to store solar heating energy not only requires a
high rectification efficiency but also high heat flow. The later
requires a high QD density in the QD junction thermal diode. The
main goal of this study is to illustrate that the multiple QDs
embedded into an insulator connected with metallic electrodes and
with a vacuum layer insert can give rise to significant thermal
rectification and negative differential thermal conductance (NDTC)
effects in the nonlinear response regime. We also clarify the
relation between the thermal power and the rectification effect.

The proposed isulator/quantum dots/vacuum (IQV) double barrier
tunnel junction system (as illustrated in Fig. 1) can be
adequately described by a multi-level Anderson model.$^{7}$ Here,
the vacuum layer serves as a blocking layer for phonon
contributions to thermal conduction, while allowing electrons to
tunnel through. We assume that the energy level separation between
the ground state and the first excited state within each QD is
much larger than $k_BT$, where $T$ is the temperature of concern.
Therefore, there are only one energy level for each QD. We have
ignored the interdot hopping terms due to the high potential
barrier separating QDs. The key effects included are the intradot
and interdot Coulomb interactions and the coupling between the QDs
with the metallic leads. Using the Keldysh-Green's function
technique,$^{8}$ the charge and heat currents through the junction
can be expressed as \bean J_e&=&\frac{-2e}{h}\sum_{\ell} \int
d\epsilon \gamma_{\ell}(\epsilon) ImG^r_{\ell,\sigma}(\epsilon)
f_{LR}(\epsilon), \eean
\begin{small}
\be Q =\frac{-2}{h}\sum_{\ell} \int d\epsilon
\gamma_{\ell}(\epsilon)
ImG^r_{\ell,\sigma}(\epsilon)(\epsilon-E_F-e\Delta V)
f_{LR}(\epsilon) , \ee
\end{small}
where $\gamma_{\ell}(\epsilon)=\frac{\Gamma_{\ell,L}(\epsilon)
\Gamma_{\ell,R}(\epsilon)}
{\Gamma_{\ell,L}(\epsilon)+\Gamma_{\ell,R}(\epsilon)}$ is the
transmission factor.
$f_{LR}(\epsilon)=f_L(\epsilon)-f_R(\epsilon)$ and
$f_{L(R)}(\epsilon)=1/(exp^{(\epsilon-\mu_{L(R)})/(k_BT_{L(R)})}+1)$
is the Fermi distribution function for the left (right) electrode.
The chemical potential difference between these two electrodes is
related to the bias difference $\mu_{L}-\mu_{R}=e \Delta V$
created by the temperature gradient. $T_L (T_R)$ denotes the
temperature maintained at the left (right) lead.
$E_F=(\mu_{L}+\mu_{R})/2$ denotes the average Fermi energy of the
electrodes. $\Gamma_{\ell,L}(\epsilon)$ and
$\Gamma_{\ell,R}(\epsilon)$ [$\Gamma_{\ell,\beta}=2\pi\sum_{{\bf
k}} |V_{\ell,\beta,{\bf k}}|^2 \delta(\epsilon-\epsilon_{{\bf
k}})]$ denote the tunneling rates from the QDs to the left and
right electrodes, respectively. $e$ and $h$ denote the electron
charge and Plank's constant, respectively. For simplicity, these
tunneling rates are assumed to be energy- and bias-independent.
Eqs. (1) and (2) have been employed to study the thermal
properties of single-level QD in the Kondo regime.$^{9}$ Here, our
analysis is devoted to the multiple-QD system in the Coulomb
blockade regime. The expression of the retarded Green function for
dot $\ell$ of a multi-QD system, $G^{r}_{\ell,\sigma}(\epsilon)$
can be found in Ref. [7]

To study the direction-dependent heat current, we let
$T_L=T_0+\Delta T/2$ and $T_R=T_0-\Delta T/2$, where
$T_0=(T_L+T_R)/2$ is the equilibrium temperature of two side
electrodes and $\Delta T=T_L-T_R$ is the temperature difference.
Because the electrochemical potential difference, $e\Delta V$
yielded by the thermal gradient could be significant, it is
important to keep track the shift of the energy level of each dot
according to $\epsilon_{\ell}=E_{\ell}+\eta_{\ell}\Delta V/2$,
where $\eta_{\ell}$ is the ratio of the distance between dot
$\ell$ and the mid plane of the QD junction to the junction width.
Here we set $\eta_B=\eta_C=0$. A functional thermal rectifier
requires a good thermal conductance for $\Delta T > 0 $, but a
poor thermal conductance for $\Delta T < 0 $. Based on Eqs. (1)
and (2), the asymmetrical behavior of heat current with respect to
$\Delta T$ requires not only highly asymmetric coupling strengthes
between the QDs and the electrodes but also strong electron
Coulomb interactions between dots. To investigate the thermal
rectification behavior, we have numerically solved Eqs. (1) and
(2) for multiple-QD junctions involving two QDs and three QDs for
various system parameters. We first determine $\Delta V$ by solving
Eq. (1) with  $J_e=0$ (the open circuit condition) for a given
$\Delta T$, $T_0$ and an initial guess of the average one-particle
and two-particle occupancy numbers, $N_{\ell}$ and $c_{\ell}$ for
each QD. Those numbers are then updated according to Eqs. (5) and
(6) in Ref. [7] until self-consistency is established. For the
open circuit, the electrochemical potential will be formed due to
charge transfer generated by the temperature gradient. This
electrochemical potential is known as the Seebeck voltage (Seebeck
effect). Once $\Delta V$ is solved, we then use Eq. (2) to compute
the heat current.

Fig. 2 shows the heat currents, occupation numbers, and
differential thermal conductance (DTC) for the two-QD case, in which the
energy levels of dot A and dot B are $E_A=E_F-\Delta E/5$ and
$E_B=E_F+\alpha_B\Delta E$, where $\alpha_B$ is tuned between 0
and 1. The heat currents are exporessed in units of $Q_0=\Gamma^2/(2h)$ through out this
article. The intradot and interdot Coulomb interactions used are
$U_{\ell}=30k_BT_0$ and $U_{AB}=15k_BT_0$. The tunneling rates are
$\Gamma_{AR}=0$, $\Gamma_{AL}=2\Gamma$, and
$\Gamma_{BR}=\Gamma_{BL}=\Gamma$. $k_BT_0$ is chosen to be
$25\Gamma$ throughout this article.  Here, $\Gamma
=(\Gamma_{AL}+\Gamma_{AR})/2$ is the average tunneling rate in
energy units, whose typical values of interest are between 0.1 and
0.5 meV. The dashed curves are obtained by using a simplified
expression of Eq. (2) in which we set the
average two particle occupation in dots A and B to zero (resulting from
the large intradot Coulomb interactions) and taking the limit
that $\Gamma \ll k_BT_0$ so the Lorentzian function of resonant
channels can be replaced by a delta function. We have
\begin{eqnarray}
{Q}/\gamma_B &=& \pi (1-N_B)[(1-2N_A)(E_B-E_F)f_{LR}(E_B) \\
\nonumber &+&2N_A(E_B+U_{AB}-E_F)f_{LR}(E_B+U_{AB})],
\end{eqnarray}
Here $N_{A(B)}$ is the average occupancy in dot A(B). Therefore,
it is expected that the curve corresponding to $E_B=E_F+4\Delta
E/5$ obtained with this delta function approximation is in good
agreement with the full solution, since $E_B$ is far away from the
Fermi energy level. For cases when $E_B$ is close to $E_F$, the
approximation is not as good, but it still gives qualitatively
correct behavior. Thus, it is convenient to use this simple
expression to illustrate the thermal rectification behavior. The
asymmetrical behavior of $N_A$ with respect to $\Delta T$ is
mainly resulted from the condition $\Gamma_{AR}=0$ and
$\Gamma_{AL}=2\Gamma$. The heat current is contributed from the
resonant channel with $\epsilon=E_B$, because the resonant channel
with $\epsilon=E_B+U_{AB}$ is too high in energy compared with
$E_F$. The sign of $Q$ is determined by $f_{LR}(E_B)$, which
indirectly depends on Coulomb interactions, tunneling rate ratio
and QD energy levels. The rectification behavior of $Q$ is
dominated by the factor $1-2N_A$, which explains why the energy
level of dot-A should be chosen below $E_F$ and the presence of
interdot Coulomb interactions is crucial. The negative sign of Q in the regime of $\Delta T < 0$
indicates that the heat current is from the right electrode to the
left electrode. We define the rectification efficiency as
$\eta_{Q}=(Q(\Delta T=30\Gamma)-|Q(\Delta T=-30\Gamma)|)/Q(\Delta
T=30\Gamma)$. We obtain $\eta_{Q}=0.86$ for $E_B=E_F+2\Delta E/5$
and 0.88 for $E_B=E_F+4\Delta E/5$. Fig. 2(c) shows DTC in units of $Q_0
k_B/\Gamma$. It is found that the rectification behavior is not
very sensitive to the variation of $E_B$. DTC is roughly linearly
proportional to $\Delta T$ in the range $-20\Gamma < k_B \Delta T
< 20 \Gamma$. In addition, we also find a small negative
differential thermal conductance (NDTC) for $E_B=E_F+4\Delta E/5$.
Similar behavior was reported in the phonon junction
system.$^{10}$

Fig. 3 shows the heat current, differential thermal conductance
and thermal power ($S=e\Delta V/k_B\Delta T$) as functions of
temperature difference $\Delta T$ for a three-QD case for various
values of $\Gamma_{AR}$, while keeping
$\Gamma_{B(C),R}=\Gamma_{B(C),L}=\Gamma$. Here, we adopt
$\eta_A=|\Gamma_{AL}-\Gamma_{AR}|/(2\Gamma)$ instead of fixing
$\eta_A$ at 0.3 to reflect the correlation of dot position with
the asymmetric tunneling rates. We assume that the three QDs are
roughly aligned with dot A in the middle. The energy levels of
dots A, B and C are chosen to be $E_A=E_F-\Delta E/5$,
$E_B=E_F+2\Delta E/5$ and $E_C=E_F+3\Delta E/5$.
$U_{AC}=U_{BA}=15k_BT_0$, $U_{BC}=8k_BT_0$, $U_C=30 k_BT_0$, and
all other parameters are kept the same as in the two-dot case. The
thermal rectification effect is most pronounced when
$\Gamma_{AR}=0.$ as seen in Fig. 4(a). (Note that the heat current
is not very sensitive to $U_{BC}$). In this case, we obtain a
small heat current $Q=0.068 Q_0$ at $\Delta T=-30\Gamma$, but a
large heat current $Q=0.33 Q_0$ at $\Delta T =30 \Gamma$ and the
rectification efficiency $\eta_{Q}$ is 0.79. However, the heat
current for $\Gamma_{AR}=0$ is small. For $\Gamma_{AR}=0.1\Gamma$,
we obtain $Q=1.69 Q_0$ at $\Delta T=-30\Gamma$,$Q=5.69 Q_0$ at
$\Delta T =30 \Gamma$, and $\eta_{Q}= 0.69$. We see that the heat
current is  suppressed for $\Delta T < 0$ with decreasing
$\Gamma_{AR}$. This implies that it is important to blockade the
heat current through dot A to observe the rectification effect.
Very clear NDTC is observed in Fig. 3(b) for the
$\Gamma_{AR}=0.1\Gamma$ case, while DTC is symmetric with respect
to $\Delta T$ for the $\Gamma_{AR}=\Gamma_{AL}$ case.

From the experimental point of view, it is easier to measure the
thermal power than the direction-dependent heat current. The
thermal power as a function of $\Delta T$ is shown in Fig. 3(c).
All curves except the dash-dotted line (which is for the
symmetrical tunneling case) show highly asymmetrical behavior with
respect to $\Delta T$, yet it is not easy at
all to judge the efficiency of the rectification effect from $S$ for small $|\Delta
T|$ ($k_B|\Delta T|/\Gamma < 10$). Thus, it is not sufficient to
determine whether a single QD can act as an efficient thermal rectifier based on
results obtained in the linear response regime of $\Delta T/T_0
\ll 1$.$^{6}$ According to the thermal power values, the
electrochemical potential $e\Delta V$ can be very large.
Consequently, the shift of QD energy levels caused by $\Delta V$
is quite important. To illustrate the importance of this effect,
we plot in Fig. 4 the heat current for various values of $E_C$ for
the case with $\Gamma_{AR}=0$, $U_{BC}=10k_BT_0$ and $\eta_A=0.3$.
Other parameters are kept the same as those for Fig. 3. The solid
(dashed) curves are obtained by including (excluding) the energy
shift $\eta_A \Delta V/2$. It is seen that the shift of QD energy
levels due to $\Delta V$ can lead to significant change in the
heat current. It is found that NDTC is accompanied with low heat current
for the case of $E_C=E_F+\Delta E/5$ [see Fig. 4(b)]. Even though
the heat currents exhibits rectification effect for $E_C=E_F+\Delta
E/5$ and $E_C=E_F+3\Delta E/5$, the thermal powers have very
different behaviors. From Figs. 3(c) and 4(c), we see that the
heat current is a highly nonlinear function of electrochemical
potential $\Delta V$. Consequently, the rectification effect is
not straightforwardly related to the thermal power in this system.

Comparing the heat currents of the three-dot case (shown in Figs. 3 and 4) to
the two-dot case (shown in Fig. 2), we find that the rectification
efficiency is about the same for both cases, while the magnitude
of the heat current can be significantly enhanced in the three-dot
case. For practical applications, we need to estimate the
magnitude of the heat current density and DTC of the IQV junction
device in order to see if the effect is significant. We envision a
thermal rectification device made of an array of multiple QDs
(e.g. three-QD cells) with a 2D density $N_{2d} = 10^{11} cm^{-2}$.
For this device, the heat current density versus $\Delta T$ is
given by Figs. 3 and 4 with the units $Q_0$ replaced by $N_{2d}Q_0$,
which is approximately $965 {} W/m^2$ if we assume $\Gamma=0.5
meV$. Similarly, the units for DTC becomes $N_{2d}k_B Q_0
/\Gamma$, which is approximately $34 {} W/{}^0Km^2$. Since the
phonon contribution can be blocked by the vacuum layer in our
design, this device should have practical applications near
$140^0K$ with ($k_BT_0 \approx 12.5 meV$). If we choose a higher tunneling rate $\Gamma > 1 meV$ and Coulomb
energy $>300 meV$ (possible for QDs with diameter less than 1 nm), then it is possible to chieve room-temperature operation.

In summary, we have reported a design of multiple-QD junction which can have significant thermal rectification effect.
The thermal rectification behavior is sensitive to the coupling between the QDs and the
electrodes, the electron Coulomb interactions and the energy
level differences between the dots.


{\bf Acknowledgments}

This work was supported by Academia Sinica, Taiwan.

Email-address:
mtkuo@ee.ncu.edu.tw; yiachang@gate.sinica.edu.tw

\mbox{}


{\bf Figure Captions}

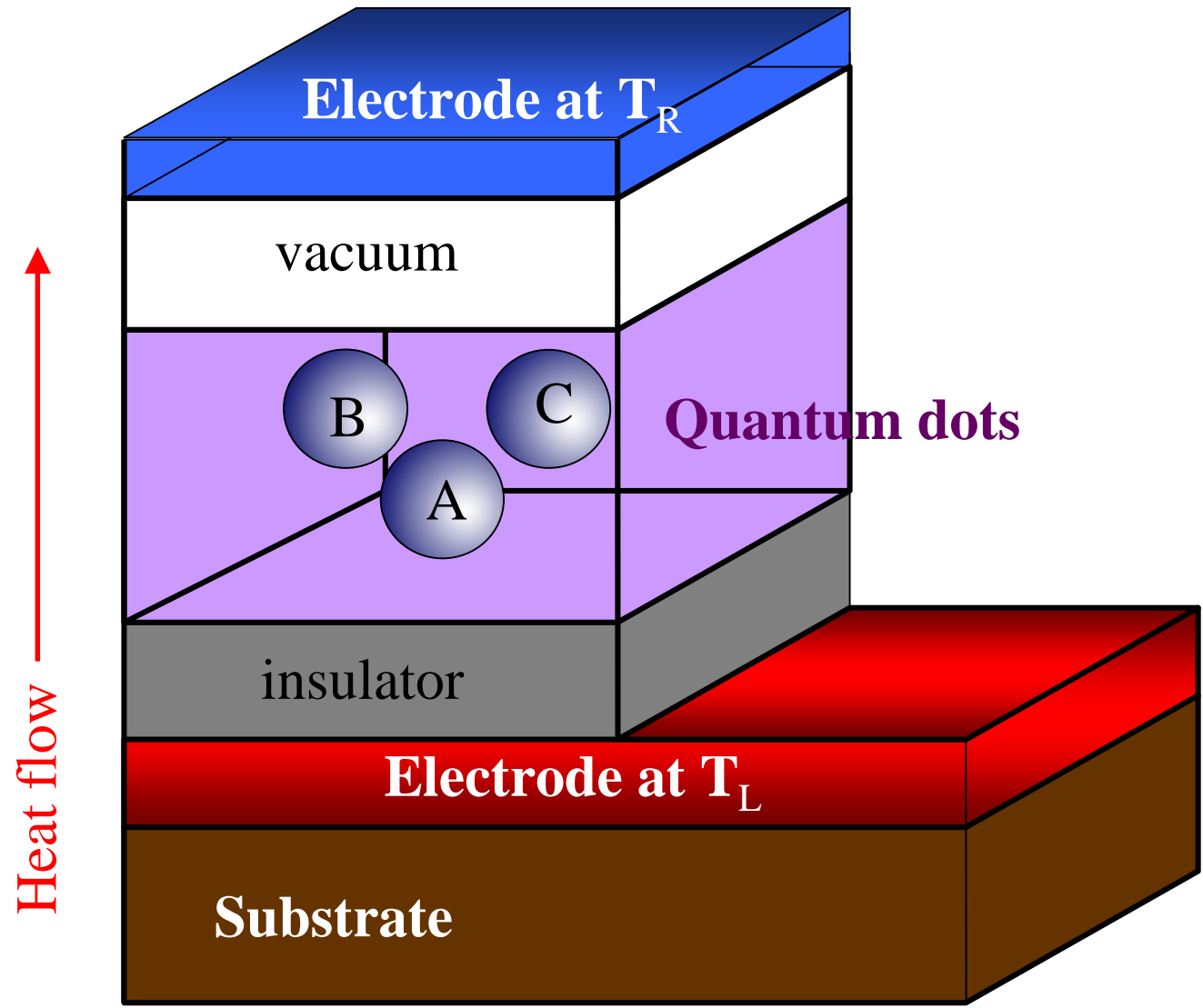
Fig. 1. Schematic diagram of the isulator/quantum dots/vacuum tunnel junction device.

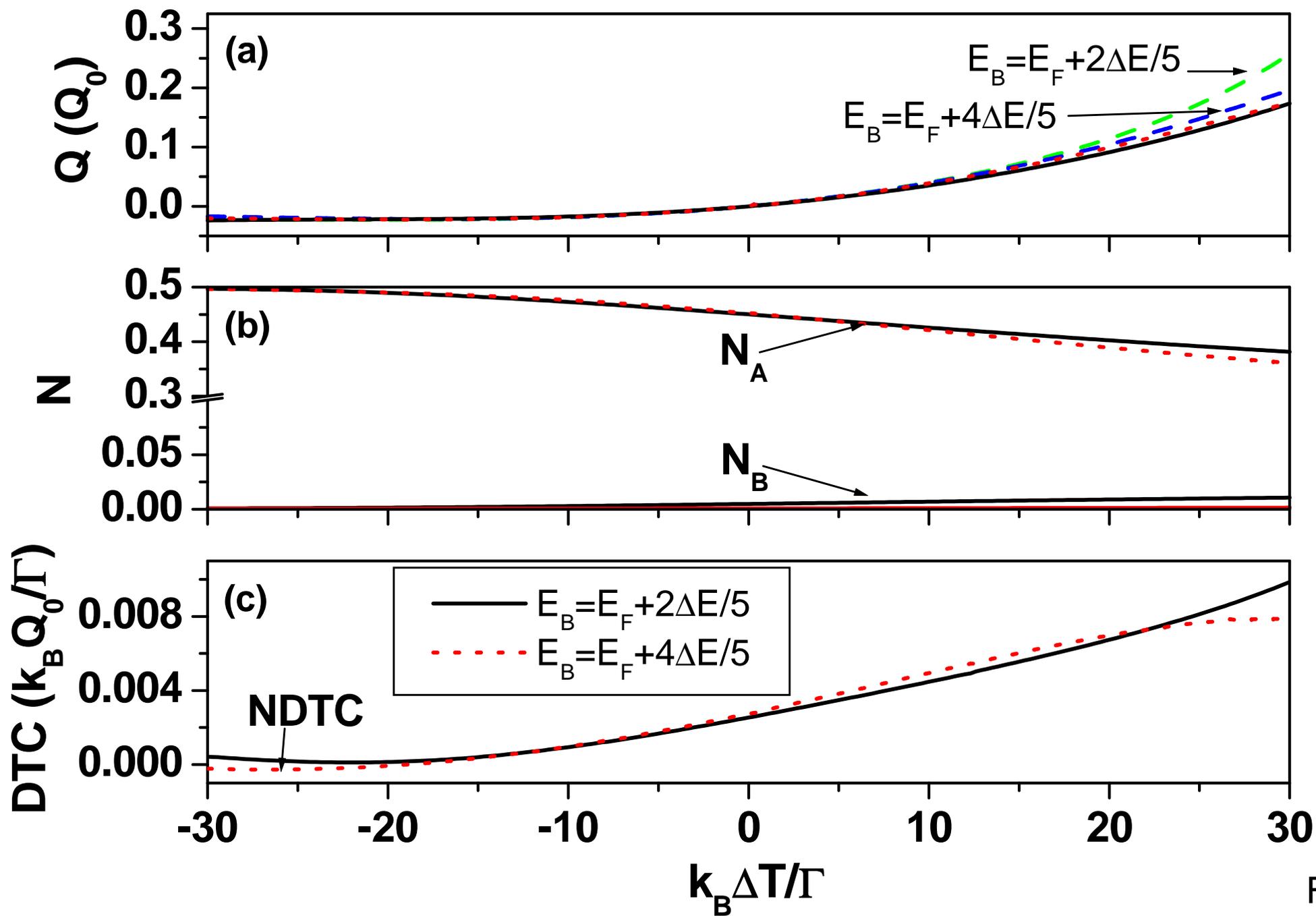
Fig. 2. (a) Heat current (b) average occupation number, and (c)
differential thermal conductance as a function of $\Delta T$ for
various values of $E_B$ for a two-QD junction.
$\Gamma_{AR}=0$, $\eta_A=0.3$ and $\Delta E= 200 \Gamma$.

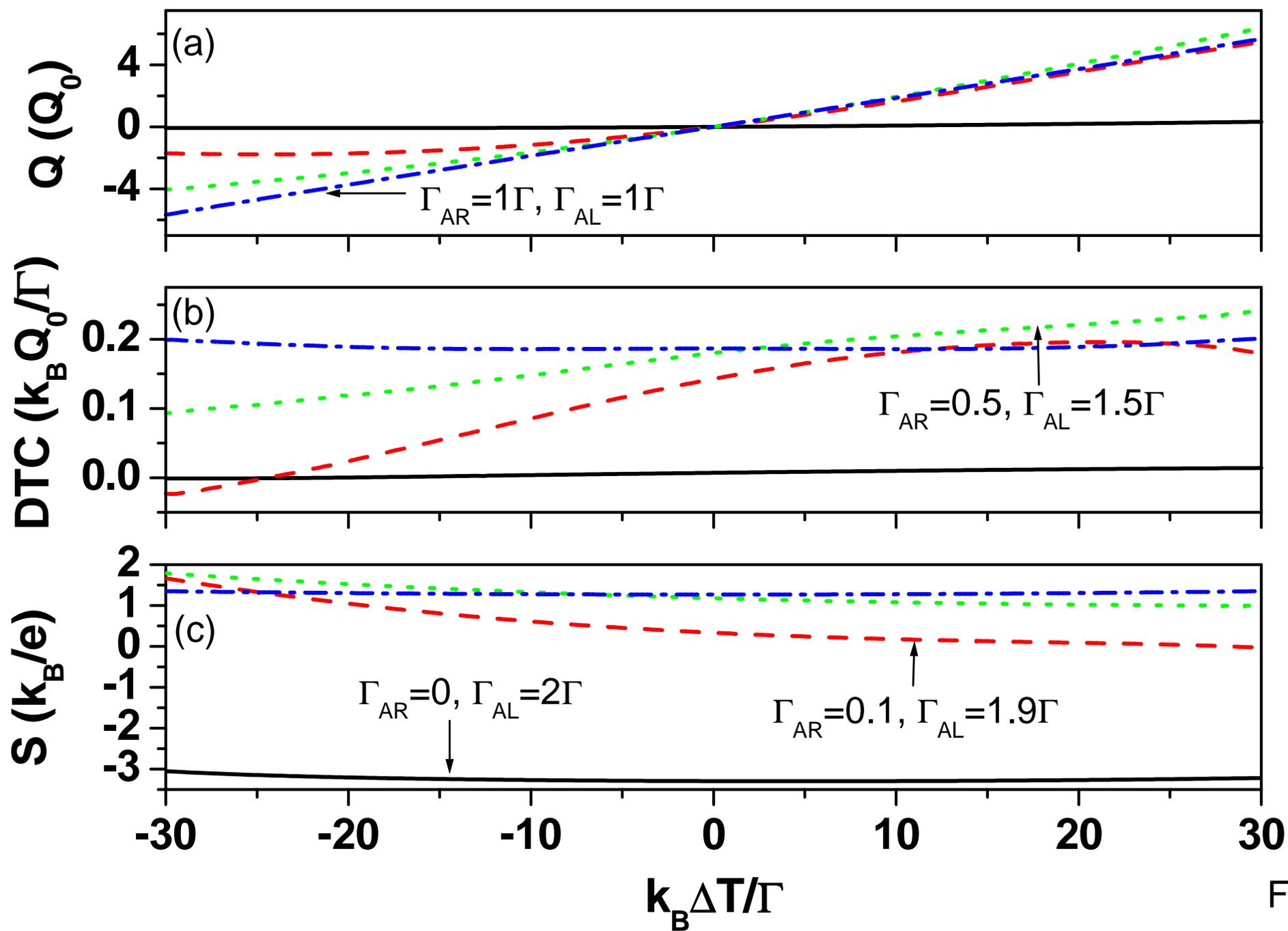
Fig. 3. (a) Heat current, (b) differential thermal conductance and
(c) thermal power as a function of $\Delta T$ for various values
of $\Gamma_{AR}$ for a three-QD junction.

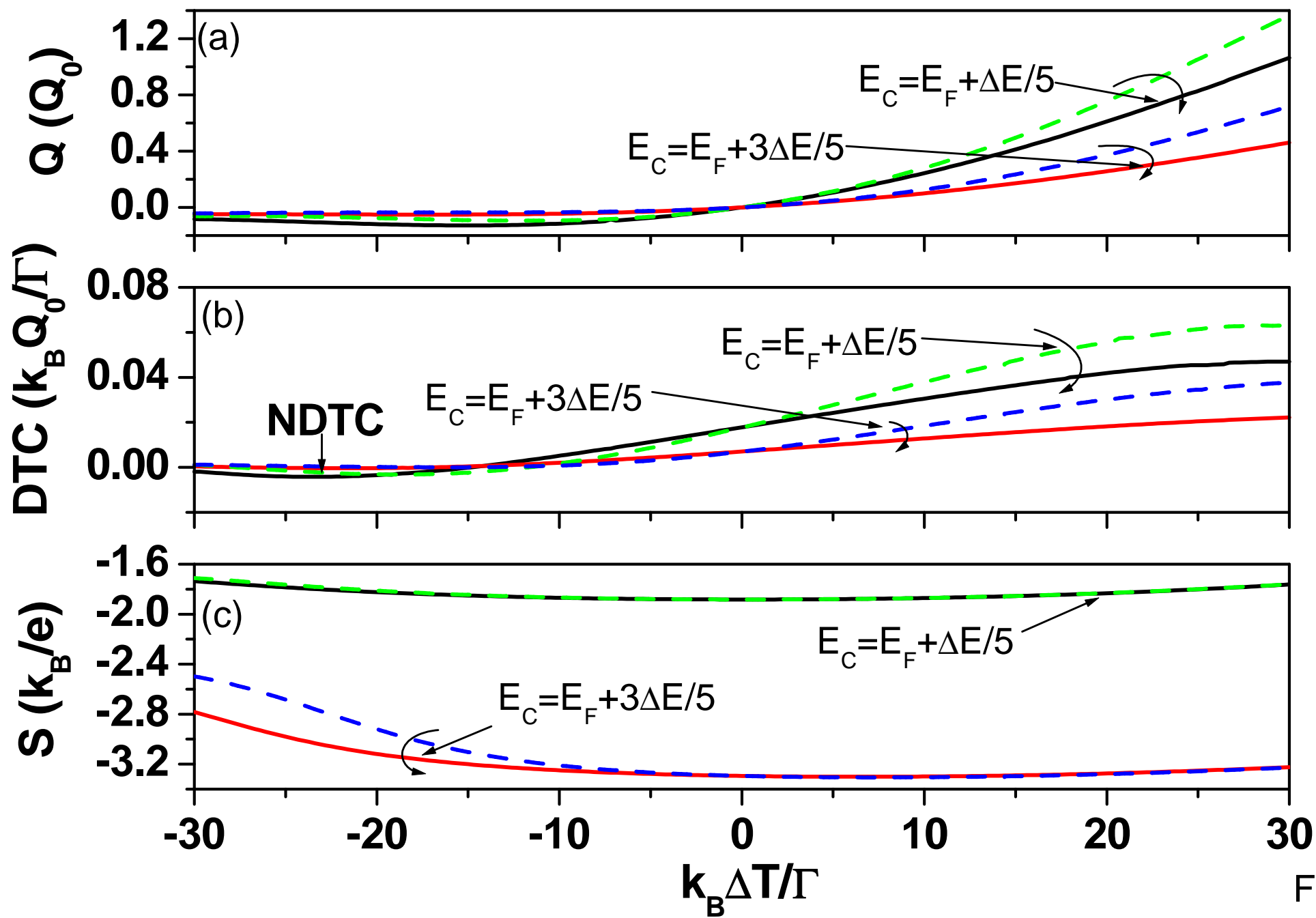
Fig. 4. (a) Heat current, (b) differential thermal conductance and
(c) thermal power as functions of $\Delta T$ for various values of
$E_C$ for a three-QD junction with $\Gamma_{AR}=0$ and
$\eta_A=0.3$.

\end{document}